\documentclass[10pt,preprint2]{aastex}

\begin{document}

\title{Fireballs, Flares and Flickering: A Semi-analytic, LTE, 
Explosive Model from Accretion Disks to Supernovae}
\author{K.\ J.\ Pearson}
\affil{Louisiana State University, Department of Physics and Astronomy,
Nicholson Hall, Baton Rouge, LA~70803-4001}
\author{Keith Horne} 
\affil{School of Physics and Astronomy, University of St.\ Andrews, 
North Haugh, St.\ Andrews KY16 9SS}
\and
\author{Warren Skidmore}
\affil{Caltech, Mail code 105-24, Pasadena, CA 91125-24}
\shorttitle{Fireballs, Flares and Flickering}
\shortauthors{K.\ J.\ Pearson, K.\ Horne and W.\ Skidmore}

\begin{abstract} 
We derive simple analytic expressions for the continuum lightcurves and 
spectra of flaring and flickering events that occur over a wide range of 
astrophysical systems. We compare these results to data taken from the 
cataclysmic variable SS Cygni and also with SN~1987A,
deriving physical parameters for the material involved. Fits to the data 
indicate a nearly time-independent photospheric temperature arising from the 
strong temperature dependence of opacity when hydrogen is partially ionized.
\end{abstract} 


\section{Introduction}
\label{sec:intro}

In a recent paper \citep{pearson03}, we explained the unusual flaring activity 
of the cataclysmic
variable (CV) system AE~Aqr in terms of the aftermath of the collision between
two gas clouds. We modelled the resulting fireball numerically, comparing
the results to analytic approximations for the optical 
lightcurves and continuum 
spectra and to observed lightcurves and spectra. We set out here an improved
analytic calculation for an expanding fireball with an LTE ionization
structure and compare our results to more detailed
numerical simulations. We fit our results to multiwavelength lightcurves of 
two very different 
systems; deriving values for the physical conditions in them.

Flickering and flaring occurs across the whole range of astrophysical systems
from stars to active galaxies. It appears to be a 
recurrent feature of accreting systems and, in particular, those where
an accretion disk is present. This flickering process may well be associated
with the anomalously high viscosity present in these systems and represent
the effect of magnetic reconnections such as that from the viscosity
mechanisms of \citet{hawley91}. Such a sudden localised deposition of energy 
would raise the gas temperature and the consequent 
overpressure would give rise to a local expansion of the disk material. 
With material cooling adiabatically, this expansion would rapidly become 
supersonic. 

Observations by \citet{bruch00} suggest that the
flickering in CVs is not uniformly distributed across the 
disk but instead is associated with the stream-disk impact point and the
innermost boundary layer region of the disk. Similarly \citet{patterson81} found that
the flickering in HT Cas was associated with the inner part of the accretion
disk. These findings, however, are contradicted by \citet{welsh95} for HT Cas, by
\citet{baptista04} for V2051 Oph and \citet{baptista02} for the Low-Mass X-ray Binary 
(LMXB) X1822-371. All these studies show flickering to arise from a range of disk radii.

The terms ``flickering'' and ``flaring'' are often used interchangably when describing
the stochastic variability of accreting sources and, even when defined in a 
particular field, are not necessarily used consistently. \citet{warner95}
and \citet{baptista04} describe flickering in CVs as the continuous, random brightness 
fluctuations of 0.01--1~mag on timescales of seconds to dozens of minutes although the 
exact numerical ranges differ between the two. In contrast the unusal CV AE~Aqr is 
universally described as a ``flaring'' source since it has rarer by brighter events that 
often occur in batches. These events can raise the total optical luminosity of the system
by factors of 2-3 contrasting with the 5-20\% typical of CV flickering \citep{bruch92}
but on a similar timescale. Consequently, we adopt the convention of describing the 
small amplitude, continuous variations as ``flickering'' and reserve the term ``flaring''
for larger scale events.

The underlying model used to reproduce the AE Aqr lightcurves was based
on consideration of the flux emitted by a hot, spherically symmetric ball of 
gas as it expanded and cooled. While we saw that the ``expanding fireball'' 
situation in
AE~Aqr could arise from collisional heating of gas blobs, the same situation
might arise from the very different causes outlined above. 
As a result, these ``fireball'' models may have a much 
wider range of applicablility than simply the unusual behaviour of AE Aqr.
Accordingly, in this paper, we will compare analytic expressions derived 
for the lightcurves of such expansions to observations of SS Cygni and 
SN~1987A.

SS Cyg is a member of the dwarf nova subclass of cataclysmic variables. It
consists of a $1.2 M_\odot$ white dwarf accreting material through an
accretion disk from a $0.71 M_\odot$ K5V secondary that loses material via 
Roche lobe overflow \citep{ritter03}. SS~Cyg
was the second member of this subclass identified \citep{wells96} and has had
a near continuously monitored lightcurve for over a century \citep{warner95}.
Like all dwarf novae, SS Cyg exhibits optical variability over a large range 
of timescales e.g. outbursts ($\Delta m\approx3.5$ mag, $t\approx40$ days), 
orbital modulations ($\Delta m\approx0.5$ mag, $P=0.275130$ days), flickering 
($\Delta m=0.01-0.2$ mag, $t\sim1$ min) and Dwarf Nova Oscillations 
($\Delta m\approx0.02$ mag, $t\approx2.8-10.9$~s) 
\citep{mauche01,teeseling97,honey89,warner95,warner04}.
We shall concentrate our modelling on the lightcurve from a flickering event.

SN~1987A was the brightest supernova observed since Kepler's in 1604. The
lightcurve can be broken into three phases: a ``flash'' lasting a few hours, 
a subsequent ``bump'' lasting around 4 months and a final exponential ``tail''
powered by radioactive decay of unstable nuclei in the ejecta.
It was a type II supernova with an identified progenitor (Sk-$69^{o}202$)
that was a blue supergiant. It is  believed that it lasted $\sim10^{7}$~y
as a 20-$30 M_{\odot}$ main sequence blue supergiant star, before 
becoming a red supergiant for $\sim10^{6}$~y during which phase it lost 
3-$6 M_{\odot}$ before finally becoming a blue supergiant again for a few
thousand years \citep{mccray93}. The rebrightening ``bump'' phases arises 
from a photosphere moving out with expelled material and eventually reversing
as the material becomes optically thin again. It is interesting to note that
the photosphere maintained a roughly constant temperature $\sim6000$~K
close to the recombination temperature for hydrogen at the expected 
densities in the expanding shell. The models of AE~Aqr flares in 
\citet{pearson03} showed a similar isothermal behaviour.

\section{Model Assumptions}
\label{sec:assume}

We briefly recap here the dynamical model developed in \citet{pearson03} but 
the interested reader is referred there for a more detailed derivation.
Let us assume a spherically symmetric expansion of a Gaussian density profile
with radial velocity 
proportional to the distance from the centre of the expansion.
We define
\begin{equation} 
\eta\equiv\frac{r}{a}
\end{equation}
as a dimensionless measure of the radius $r$
in terms of the current lengthscale of the Gaussian $a$. The expansion 
factor $\beta$ acts as a dimensionless time, being the constant
of proportionality between the current and fiducial
scalelength $a_{0}$. Hence,
\begin{equation}
\beta\equiv\frac{a}{a_{0}}
\end{equation} 
consistent with our above assumptions regarding velocity and implying
\begin{equation}
\beta=1+H\ (t-t_{0})
\label{eqn:betaH}
\end{equation}
where $t_0$ is the time at  which all the fiducial values are determined and
$H$ is an ``expansion constant'' setting the speed of the expansion.
For simplicity, we
consider only the case of uniform 3-D expansion, avoiding factors such 
as the angle of the observer to lines of symmetry. We also restrict ourselves
to a spatially uniform temperature distribution since, for example,
a power law distribution gives a result for the later integration in
(\ref{eqn:tauint}) in terms of the hypergeometric function $_{2}F_{1}$. 
We assume a power law with index $\Gamma$ for the temporal dependance of 
temperature. $\Gamma=0$ corresponds to an isothermal expansion and 
$\Gamma=2$ to the adiabatic case.    

In summary then, we  have
\begin{eqnarray}
T & = & T_{0}\ \beta^{-\Gamma}, \label{eqn:tempdep}\\
\rho & = & \rho_{0}\ \beta^{-3}\ e^{-\eta^{2}}, \label{eqn:densprof}
\end{eqnarray}
where
\begin{equation}
\rho_{0} = \frac{M}{\left(\pi\ a_{0}^{2}\right)^{\frac{3}{2}}}
\end{equation}
and $M$ is the total mass involved in the expansion.

Using $v(r)=H r_{0}=H r \beta^{-1}$ we can integrate $\frac{1}{2} \rho v^{2}$
over all space to derive the total kinetic energy of the expansion
\begin{eqnarray}
E_{\rm kin} & = &\frac{2}{\sqrt{\pi}}\ M\ a_{0}^{2}\ H^{2} 
\int_0^\infty \eta^{4}\ e^{-\eta^{2}}\ {\rm d}\eta \\
 & = & \frac{3}{4}\ M\ a_{0}^{2}\ H^{2} \\
 & \equiv & \frac{3}{4}\ M\ v_{0}^{2} 
\end{eqnarray}
defining $v_{0}=H a_{0}$, the speed of expansion at $r=a$.

\section{Theoretical Lightcurves and Spectral Distributions}
\label{sec:derive}

The radiative transfer equation has a formal solution under conditions of LTE
\begin{equation}
	I = \int_{0}^{\infty} B\ e^{-\tau}\ {\rm d}\tau , \label{radtrans}
\end{equation}
where $I$ is the intensity of the emerging radiation,
$B$ is the Planck function and $\tau$ is the optical depth
measured along the line of sight from the observer. For cases where the source 
function is everywhere the same this becomes
\begin{equation}
I=B\ (1-e^{-\tau}) .\label{radtran2}
\end{equation}

We define $x$ as the distance from the fireball centre toward the
observer, and $y$ as the distance perpendicular to the line of sight.
The above line integral (equation (\ref{radtran2})) gives the intensity
$I(y)$ for lines of sight with different impact parameters $y$.
The fireball flux, obtained by summing intensities
weighted by the solid angles of annuli on the sky, is then
\begin{equation}
f(\lambda) = \int_0^\infty I(y) \frac{2\pi\ y}{ d^2 }\ {\rm d}y 
\label{eqn:flyint}
\end{equation}
where $d$ is the source distance.

We can calculate the evolution of the continuum lightcurve using  
expressions for the linear absorption coefficient. The free-free absorption 
coefficient (per unit distance) can be written as
\begin{equation}
\kappa_{\rm ff}=\kappa_0\ \left[1-e^{-\left(\frac{h\nu}{kT}\right)}\right] 
\ T^{-\frac{1}{2}}\ \nu^{-3}\ n_{\rm e}\ n_{\rm i}
\label{eqn:ffopac}
\end{equation}
where
\begin{equation}
\kappa_0 = 3.692\times10^{-2}\ Q^{2}\ g_{\rm ff}~{\rm (SI)}
\end{equation}
\citep{keady00}, $g_{\rm ff}$ is the free-free Gaunt factor \citep{gaunt30} 
and $Q$ the ionic charge. We wish to retain the explicit frequency dependence 
but otherwise follow a parallel derivation as for Kramers' opacity. For a 
mixed elemental composition, then, we sum over all species (assumed fully 
ionized)
\begin{equation}
\sum_{\rm all~ions} Q_{\rm i}^{2}\ n_{\rm i}\ g_{\rm i,ff} \approx 
\frac{\rho}{m_{\rm H}}\ (1-Z)\ \bar{g}_{\rm ff} 
\end{equation}
\citep{bowers84}. Here $\bar{g}_{\rm ff}$ is a mean Gaunt factor (close to 
unity) and $Z$ the metal mass fraction. We also have
\begin{equation}
n_{\rm e} = \frac{\rho}{\mu_{\rm e}\ m_{\rm H}} = 
\frac{\rho\ (1+X)}{2\ m_{\rm H}}
\end{equation}
\citep{bowers84} where $X$ is the hydrogen mass fraction. Combining and 
writing the correction for stimulated emission as 
\begin{equation}
\epsilon=1-e^{-\left(\frac{h\nu}{kT}\right)}
\end{equation}
gives
\begin{equation}
\kappa_{\rm ff}=\kappa_1\ \epsilon\ \rho^{2}\ T^{-\frac{1}{2}}\ \nu^{-3} 
\label{eqn:ffopac2}
\end{equation} 
where we have defined
\begin{equation}
\kappa_{1} = 6.695\times10^{51}\ (1-Z)\ (1+X) 
\bar{g}_{\rm ff}~{\rm (SI)}. 
\label{eqn:kap1ff}
\end{equation}
Inserting our density profile (equation  (\ref{eqn:densprof})) we arrive at 
the result
\begin{equation}
\kappa_{\rm ff}=\frac{\kappa_1\ \epsilon}{T^{\frac{1}{2}}\ \nu^{3}} 
\frac{M^{2} }{\pi^{3}\ a^{6}}\ e^{-2\eta^{2}}.
\label{eqn:ffmopac}
\end{equation}

On dimensional grounds if on no other, a similar expression to 
equation (\ref{eqn:ffopac2}) and hence (\ref{eqn:ffmopac})
must also exist for bound-free opacity when it dominates (when most 
species are fully recombined). Textbook derivations
of Kramer's opacity (eg. Bowers \& Deeming 1984) show us
\begin{equation}
\kappa_{1,{\rm bf}} \propto Z\ (1+X)\ \bar{g}_{\rm bf}
\label{eqn:kap1bf}
\end{equation}
where $g_{\rm bf}$ is the bound-free Gaunt factor \citep{gaunt30}. Useful tables
for $g_{\rm bf}$ have been calculated by \citet{glasco64}.
The constant of proportionality, however, is more difficult to determine than 
for free-free, since it must account
for the ionization edges in the absorption and, in particular, the 
change of energy level populations with temperature for each ion. We shall
assume a relation analagous to equation (\ref{eqn:ffopac2}) also holds for 
the situation of 
mixed ionized and recombined species where both forms of opacity contribute.

The optical depth parallel to the observers line of sight
\begin{eqnarray}
\tau(y) & = & - \int_{\infty}^{-\infty} \kappa\ {\rm d}x \label{eqn:tauint}\\
& = & \left[\frac{\kappa_1\ \epsilon}{T^{\frac{1}{2}}\ \nu^{3}}
\frac{M^{2} }{\pi^{3}\ a^{5}}\right]\ e^{-2(\frac{y}{a})^{2}}
\int_{-\infty}^{\infty} e^{-2(\frac{x}{a})^{2}}
\ {\rm d}\left(\frac{x}{a}\right) \nonumber \\
& &
\label{eqn:spatint}\\
&=& \tau_{0}\ e^{-2(\frac{y}{a})^{2}} \label{eqn:tauy2}
\end{eqnarray}
where the optical depth on the line of sight through the centre of the fireball
is
\begin{equation}
\tau_{0}=
\left(\frac{\beta_{\rm c}}{\beta}\right)^{\left(\frac{10-\Gamma}{2}\right)}
\label{eqn:tau0def}
\end{equation}
and the time at which the fireball becomes optically thin along this line
of sight is
\begin{equation}
\beta_{\rm c}\equiv\left[\frac{\kappa_1\ \epsilon}{T_0^{\frac{1}{2}}\ \nu^{3}}
\frac{M^{2}}{\sqrt{2\ \pi^{5}}\ a_{0}^{5}}
\right]^{\frac{2}{10-\Gamma}}. \label{eqn:beta0def}
\end{equation}
It should be noted that in pulling $\kappa_1(\mu_{\rm i},\mu_{\rm e})$ out of 
the integral (\ref{eqn:spatint}) we have implicitly assumed that the spatial 
variation of the 
ionization fraction has negligible impact on the behaviour. This is an 
assumption that we shall return to later.

\begin{figure}
\includegraphics[angle=-90,scale=0.3]{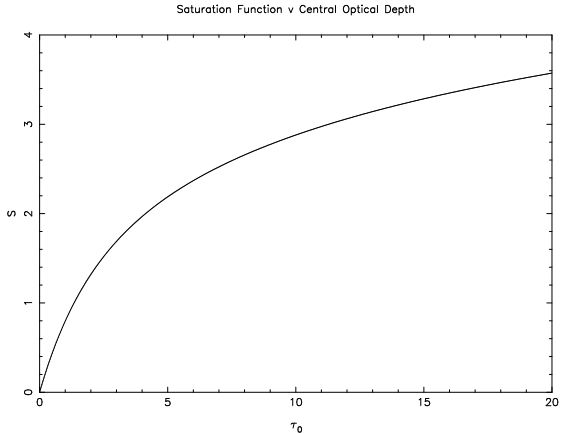}
\caption{The Saturation Function $S$.}
\label{fig:satfn}
\end{figure}

\begin{figure}
\includegraphics[angle=-90,scale=0.3]{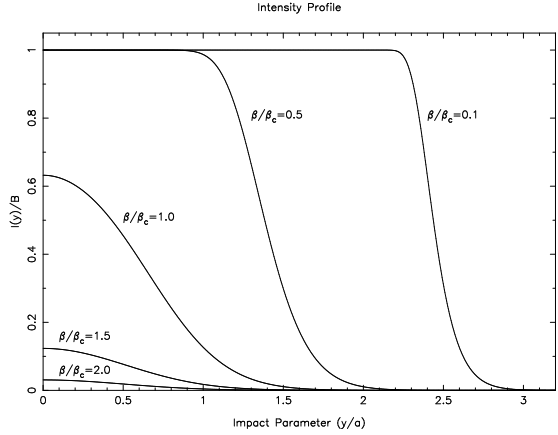}
\caption{Comparison of the intensity profiles at different times calculated 
using (\protect\ref{eqn:tauy2}) and (\protect\ref{radtran2}) for a $\Gamma=0$
(isothermal) evolution.}
\label{fig:iofyplot}
\end{figure}

In \citet{pearson03} we approximated the flux received 
from the fireball by splitting it into two components as viewed on the
plane of the sky: an optically thick central region bounded by $y=y_{\rm m}$
(see equation \ref{eqn:ymax}) and an optically thin 
surrounding. Here, however, we calculate the integral exactly. 
Combining equations (\ref{radtran2}), (\ref{eqn:flyint}) and (\ref{eqn:tauy2})
with a change of integration variable to $u=\tau(y)$, we have
\begin{eqnarray}
f & = & \frac{\pi\ a^{2}\ B}{2\ d^{2}}\int_{0}^{\tau_{0}} 
\frac{1-e^{-u}}{u}\ {\rm d}u
\end{eqnarray}
Expressing the exponential in its series form we integrate to find
\begin{equation}
f=\frac{\pi\ a^{2}\ B}{2\ d^{2}} \sum_{n=1}^{\infty} \frac{(-1)^{(n+1)}
\ \tau_{0}^{n}} {n\ n!}
\label{eqn:intseries}
\end{equation}
which can also be rewritten in the form of standard functions
\begin{equation}
f=\frac{\pi\ a^{2}\ B}{2\ d^{2}} (E_{1}(\tau_{0}) +\gamma+\ln(\tau_{0}))
\label{eqn:fluxtot}
\end{equation}
where $E_{1}$ is the first order exponential integral and 
$\gamma\approx0.577216$ is Euler's gamma constant 
\citep{abramowitz72,jeffreys56,numrec}. This has the form of
\begin{equation}
f=\Omega\ B\ S(\tau_{0}),
\end{equation}
where $\Omega=\frac{\pi a^{2}}{2 d^{2}}$ is the solid angle subtended
by the current standard deviation ($a/\sqrt{2}$) of the Gaussian density 
profile and the ``Saturation Function'' 
\begin{equation}
S(\tau_{0})=(E_{1}(\tau_{0}) +\gamma+\ln(\tau_{0}))
\end{equation}
is plotted in Figure~\ref{fig:satfn}.
We note the asymptotic limits
\begin{equation}
S(\tau) \approx \left\{ 
\begin{array}{l} 
\tau~~~~~~~~~~~~~~~~~~~~~~~~\!\mbox{for}~\tau\ll1\\
\gamma+\ln{\tau}~~~~~~~~~~~~~~~\mbox{for}~\tau\gg1.
\end{array}
\right.
\end{equation}
The intensity of emitted radiation as a function of impact parameter is
plotted for different times in Figure~\ref{fig:iofyplot}. We can see how the 
flux saturates at the black body function for highly optically thick 
impact parameters. In comoving coordinates this region gradually shrinks and 
at later times the entire emission becomes optically thin.

\begin{figure}
\includegraphics[angle=-90,scale=0.3]{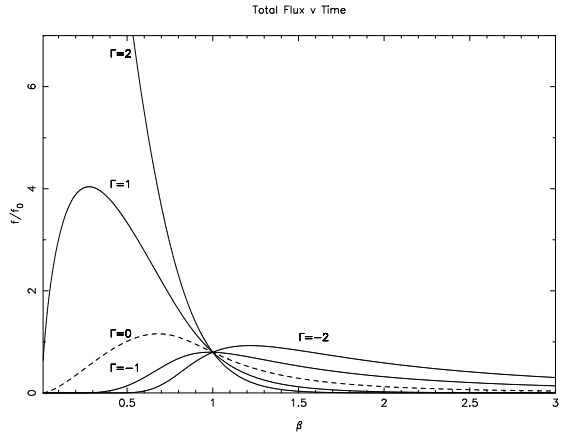}
\caption{The temporal behaviour of the total flux at $5000$~\AA ~for 
different 
cooling indices $\Gamma=-2,-1,0,1,2$, assuming $\beta_{\rm c}=1$ and 
$T_{0}=15~000$~K. The y-axis is parameterised in multiples of 
$f_{0}=\frac{\pi a_{0}}{2 d^{2}} B_{\nu}(5000~{\rm \AA}, 15~000~{\rm K})$.}
\label{fig:fluxtot}
\end{figure}

\begin{figure}
\includegraphics[angle=-90,scale=0.3]{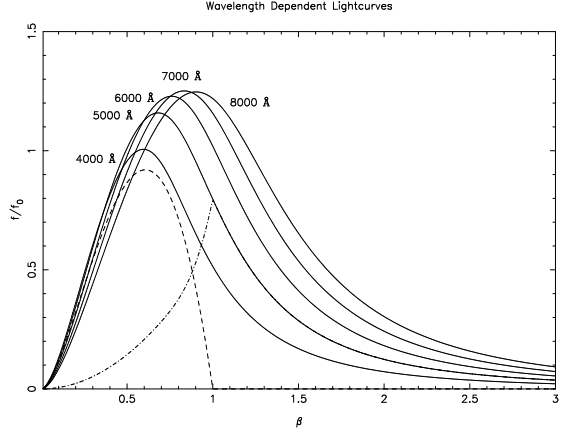}
\caption{The temporal behaviour of the total flux at several wavelengths 
assuming 
$\beta_{\rm c}=\left(\frac{\lambda}{5000~{\rm \AA}}\right)^{\frac{3}{5}}$, 
$T_{0}=15~000$~K and $\Gamma=0$. The y-axis is again parameterised in 
multiples of $f_{0}$. The optically thick contribution
at $5000$~\AA ~is plotted as a dashed line and the optically thin 
contribution as a dot-dashed line.}
\label{fig:fluxwave}
\end{figure}

The total flux given by equation (\ref{eqn:fluxtot}) is plotted against
time in Figure~\ref{fig:fluxtot} for different values of $\Gamma$, 
using $\beta_{\rm c}=1$. Figure~\ref{fig:fluxwave} shows lightcurves
at different wavelengths assuming $\Gamma=0$ (isothermal) and
$\beta_{\rm c}=(\lambda/5000~{\rm \AA})^{\frac{3}{5}}$. The 
effect of the expanding photosphere can be seen from the optically thick 
contribution to the total flux, the latter of which rises to a peak at 
$\beta_{\rm pk}$. Initially the optically thick flux is the
dominant source and the emission rises as the emitting area grows while the
photosphere is advected outwards. Eventually, the decreasing density causes
the photosphere to reverse and begin to shrink, reaching zero size at
$\beta=\beta_{\rm c}$. The 
remaining emission comes from the optically thin region surrounding
the optically thick circle as seen on the plane of the sky. This 
optically thin emission dominates at late times. 


\begin{figure}
\includegraphics[angle=-90,scale=0.3]{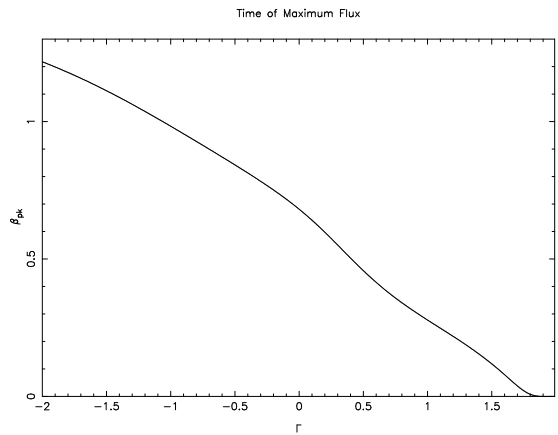}
\caption{The time of peak flux at 5000~\AA ~plotted against $\Gamma$, 
assuming $\beta_{\rm c}=1$ and $T_{0}=15~000$~K.}
\label{fig:fluxmax}
\end{figure}

\begin{figure}
\includegraphics[angle=-90,scale=0.3]{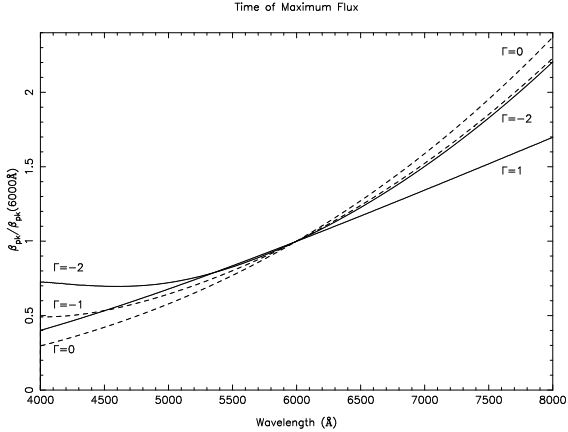}
\caption{The time of peak flux plotted against wavelength for various values of  
$\Gamma$ assuming $T_{0}=15~000$~K. The values are normalised to the peak time 
at 6000~\AA ~in each case.}
\label{fig:fluxlag}
\end{figure}

Recalling that $a=\beta a_{0}$, for the special case of $\Gamma=0$ 
(isothermal expansion) we can neglect the time variation of the 
Planck function and differentiate (\ref{eqn:fluxtot}) with respect to 
$\beta$ to find a condition for the maximum flux
\begin{equation}
4(E_{1}(\tau_{0}) +\gamma+\ln(\tau_{0}))-
      (10-\Gamma)(1-e^{-\tau_{0}})=0.
\label{eqn:turn}
\end{equation}
This can be solved numerically to find $\tau_{0{\rm,pk}}=6.8204$. From 
(\ref{eqn:tau0def}) we then find  
$\beta_{\rm pk}=0.6811\beta_{\rm c}$. We can also differentiate 
(\ref{eqn:fluxtot}) with respect to $\beta$ for $\Gamma\neq0$ although 
the form is much more untidy and we must solve for $\beta$ directly. We plot 
$\beta_{\rm pk}$ against $\Gamma$ in Figure~\ref{fig:fluxmax} continuing to 
neglect any time dependence of $\epsilon$. We plot $\beta_{\rm pk}$ against 
wavelength in Figure~\ref{fig:fluxlag} for several values of $\Gamma$.



\begin{figure}
\includegraphics[angle=-90,scale=0.3]{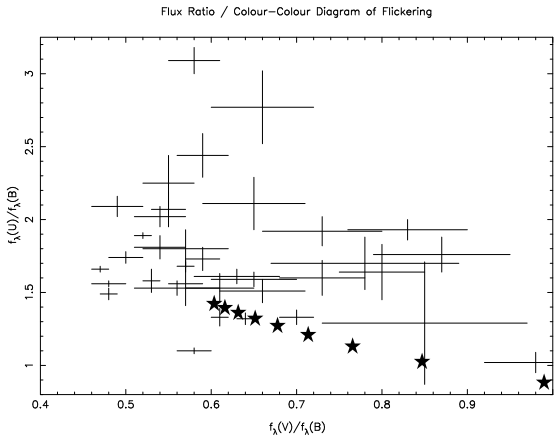}
\caption{The amplitude ratio (a U-B v V-B color-color diagram) for 
flickering in 
several CVs observed by \protect\citet{bruch92}. The predicted values are 
plotted as stars for different temperatures ranging from 8000~K
(extreme right) to 24~000~K in 2000~K steps.}
\label{fig:ampcomp}
\end{figure}

\citet{bruch92} found a correlation between the amplitude
of flickering in different wavebands for several CVs 
(see {\it inter alia} his Figure 2). Such a correlation could arise from an
isothermal evolution at a consistent temperature. This can be understood from 
(\ref{eqn:fluxtot}). Since the 
peak flux occurs at the same central optical depth independent of 
wavelength, then for two wavelengths
$\lambda_{\rm 1}$ and $\lambda_{\rm 2}$ the peak fluxes are related by 
\begin{eqnarray}
\frac{f_{\rm pk,1}}{f_{\rm  pk,2}} & = &
\left(\frac{\beta_{\rm pk,1}}{\beta_{\rm pk,2}}\right)^{2}
\frac{B_{\lambda,1}}{B_{\lambda,2}} \\
& =&
\left(\frac{\beta_{\rm c,1}}{\beta_{\rm c,2}}\right)^{2}
\left(\frac{\lambda_{1}}{\lambda_{2}}\right)^{-5} 
\left(\frac{e^{\frac{c_{2}}{\lambda_{2} T_{0}}}-1}
{e^{\frac{c_{2}}{\lambda_{1} T_{0}}}-1}\right)\\
& = &
\left(\frac{\lambda_{\rm 2}}{\lambda_{\rm 1}}\right)^{\frac{19}{5}}
\left(\frac{1-e^{\frac{-c_{2}}{\lambda_{1} T_{0}}}}
{1-e^{\frac{-c_{2}}{\lambda_{2} T_{0}}}}\right)^{\frac{2}{5}}
\left(\frac{e^{\frac{c_{2}}{\lambda_{2} T_{0}}}-1}
{e^{\frac{c_{2}}{\lambda_{1} T_{0}}}-1}\right) \nonumber \\
& &
\label{eqn:fluxrat}
\end{eqnarray}
where $c_{2}=hc/k$ and assuming no complications such as a Balmer edge between
them. 
We plot the predicted values for different temperatures 
as stars; alongside the data taken from Table~4 of \citet{bruch92},
in a color-color diagram in Figure~\ref{fig:ampcomp} (nb. we used 
$B_{\lambda}$ rather than 
$B_{\nu}$ in the above derivation purely for comparison with this dataset).


This figure shows that a simple application of our model enables us to reproduce
the observed range of V-B reasonably well using T as a free paramemter. The U-B
colour however, is generally underpredicted. This results from not making
allowance for the increased opacity above the Balmer jump when evaluating 
$\beta_{c}$ in deriving \ref{eqn:fluxrat}. The size of the Balmer jump depends upon
a non-trivial combination of mass, lengthscale and temperature to produce an effective
value of $\kappa_{1}$. Any individual system may produce flickers with a range of
physical parameters causing it to produce points scattered across this diagram.

\begin{figure}
\includegraphics[angle=-90,scale=0.3]{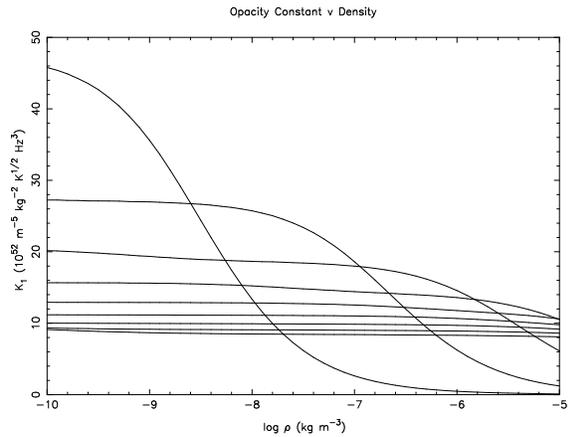}
\caption{The value of $\kappa_{1}$ at $5000$~\AA ~for temperatures ranging in 
$2000$~K steps from $8000$~K (upper left curve) to $24~000$~K.}
\label{fig:opcont}
\end{figure}

Using numerical methods outlined later, we derived values for $\kappa_{1}$ 
for several temperatures over a range of densities. These are plotted
in Figure~\ref{fig:opcont}. We can see how, for a large range of temperatures
of interest and for densities ranging over several orders of magnitude, 
$\kappa_{1}$ is well represented by a constant value. At lower temperatures
the free-free opacity transitions to bound-free opacity at lower densities
than it does at higher temperatures. At these lower temperatures we could 
introduce
significant errors if we estimate $\kappa_{1}$ from a position that has
one form of opacity when the other is in fact dominant.  
It is very unlikely that this would be
the case, however, if we estimate the ionization state from conditions at a 
suitable position and consider how $\kappa_{1}$ smoothly changes with
the ionization. Qualitatively then, at the very least,
 we can be confident of
the ability of the analysis to allow us to examine the behaviour of the 
proposed mechanism and to predict fluxes to within factors of order unity.


\section{Analysis}
\label{sec:analysis}

Equation~(\ref{eqn:fluxtot}) gives us the complete temporal and frequency
behaviour for the emergent flux in terms of $\tau_{0}$. We must use the 
Planck function appropriate to the time-dependent behaviour of the 
temperature from (\ref{eqn:tempdep}). Equation~(\ref{eqn:tau0def}) gives 
us $\tau_{0}$ as a function of $\beta$ and $\beta_{\rm c}$. 
The behaviour of the flux thus rests on the character of the parameter 
$\beta_{\rm c}$. We know functionally that
\begin{eqnarray}
\beta_{\rm c} & = & \beta_{\rm c}(\kappa_{1},\epsilon,\nu^{-3}),\\
\kappa_{1} & = &\kappa_{1}(\eta,\nu,\beta),\\
\epsilon & = & \epsilon(\nu,\beta)
\end{eqnarray}
and we shall examine each of these dependencies and appropriate levels of 
approximation in each case.

\subsection{Global Behaviour $\beta_{\rm c}={\rm const}$}

In examining the behaviour of the flux as a function of time it is instructive
to restrict ourselves initially to a single wavelength and ignore spatial
or time dependency in $\beta_{\rm c}$. 

We see from equations (\ref{eqn:tau0def}) and (\ref{eqn:intseries}) that the 
late-time behaviour ($kT\ll h\nu$) of 
the lightcurve is determined by the temperature index $\Gamma$ such that
$f\sim\beta^{-5}$ or $\beta^{-4}$ for isothermal or adiabatic cooling 
respectively.

\subsection{Spatial Dependence $\beta_{\rm c}=\beta_{\rm c}(\eta)$}

\subsubsection{One Zone Model}

We know $\beta_{\rm c}$ depends on $\kappa_{1}$ which in turn depends on the 
ionization structure across the expansion. At the outset we acknowledge
that a detailed solution accounting for the changing ionization fraction 
across the density profile would require numerical integration such
as that reported in \citet{pearson03} for AE Aqr. We are looking for simpler,
more easily calculable approximations that avoid such detailed methods.
To this end we do not include here any explicit allowance for the 
variation of ionization states across the profile; instead we approximate
the behaviour by calculating the conditions at a suitable point
within the expansion. With a Gaussian density profile we might expect
the flux to be dominated by the region close to the photosphere. As a result
we could approximate the ionization structure by the solution 
to the Saha equation under the conditions prevailing there. 

To find the position of the observed photosphere we need to integrate along
a given line of sight until optical depth unity. From our earlier
derivation we can see the optical depth down to any height $x'$ is
\begin{equation}
\tau(y)=\tau_{0}
\frac{\sqrt{2}}{\sqrt{\pi}}\ e^{-2(\frac{y}{a})^{2}} 
\int_{\frac{x'}{a}}^\infty e^{-2(\frac{x}{a})^{2}}
\ {\rm d}\left(\frac{x}{a}\right)
\end{equation}
which, setting $\tau(y)=1$ for the special case $x'$=$x_{\rm ph}$ gives
\begin{equation}
e^{2(\frac{y}{a})^{2}}=\frac{1}{2} 
\tau_{0}
\ \mbox{erfc}\left(\sqrt{2}\frac{x_{\rm ph}}{a}\right).
\end{equation}
This defines the locus of points on the photospheric surface. Since the largest
contribution to the opacity integral comes from the region of highest density,
we will introduce least error by evaluating $\kappa_{1}$ there. This will
clearly occur along the central line of sight $y=0$ and hence we arrive at
\begin{equation}
\mbox{erfc}(\sqrt{2}\eta_{\rm ph})=\frac{2}{\tau_{0}}
\label{eqn:photeval}
\end{equation}
which we must solve numerically.
However, we must remember that it is possible for the photospheric
surface to lie behind the density peak. This will just occur when
$\tau(0)=1$ at $\eta=0$, implying 
$\beta=2^{\left(\frac{2}{10-\Gamma}\right)}\beta_{\rm c}$. 
Ultimately then, we have 
\begin{equation}
\rho_{\rm eval} = \left\{ 
\begin{array}{l} 
\rho_{0}\ \beta^{-3}\ e^{-\eta_{\rm ph}^{2}}~~~~~~~\mbox{for}~\beta<2^{\left(\frac{2}{10-\Gamma}\right)}\beta_{\rm c}
\\
\rho_{0}\ \beta^{-3}~~~~~~~~~~~~~~~\mbox{otherwise}.
\label{eqn:rhoph}
\end{array}
\right.
\end{equation}

It must be emphasized that we are using this position solely to evaluate a
typical ionization state and, hence, to find a suitable approximate value for
$\kappa_{1}$. The effect of the density profile on the optical 
depth is accounted for explicitly in the derivations in 
section~\ref{sec:derive}.

However, an alternative `typical' place
to estimate $\kappa_{1}$ is at the point of maximum density ($x=0$) 
along the limiting impact parameter $y_{\rm m}$ that remains optically thick.
We can derive $y_{\rm m}$ by setting (\ref{eqn:tauy2}) equal to unity, giving 
\begin{equation}
y_{\rm m} = a_{0}\ \beta \left( \frac{\ln\tau_{0}}{2} 
\right)^{\frac{1}{2}}
\label{eqn:ymax}
\end{equation}
where $\tau_{0}>1$ ie. for times $\beta<\beta_{\rm c}$. Hence,
\begin{equation}
\rho_{\rm eval} = \left\{ 
\begin{array}{l} 
\rho_{0} 
\ \beta^{\frac{(-2-\Gamma)}{4}}\beta_{\rm c}^{\frac{(10-\Gamma)}{4}}~~~~~~~\mbox{for}~\beta<\beta_{\rm c}
\\
\rho_{0}\ \beta^{-3}~~~~~~~~~~~~~~~~~~~~~~\mbox{otherwise}.
\label{eqn:rhoym}
\end{array}
\right.
\end{equation}

\subsubsection{Pure hydrogen case - semi-analytic solution}
For simplicity, let us consider a fireball with almost pure hydrogen 
composition (nb. if $Z=0$ then equation (\ref{eqn:kap1bf}) implies 
$\kappa_{1,{\rm bf}}=0$), with an ionization fraction
\begin{equation}
\iota\equiv\frac{n_{\rm i}}{n_{\rm i}+n_{\rm n}}=\frac{n_{\rm i}}{n}.
\end{equation} 
Ignoring any contribution to opacity from H$^{\rm -}$ and noting 
$\kappa\propto\rho^{2}$, the total opacity coefficient becomes
\begin{equation}
\kappa_{1}\approx \iota^{2}\ \kappa_{1,{\rm ff}} + 
(1-\iota)^{2}\ \kappa_{1,{\rm bf}}.
\label{eqn:kap1mixcomp}
\end{equation}

For hydrogen, the Saha equation can be approximated by
\begin{eqnarray}
\frac{n_{\rm i}\ n_{\rm e}}{n_{\rm n}} 
& \approx & 2.4\times10^{21}\ T^{\frac{3}{2}}
\ \exp(-1.58\times10^{5}\ T^{-1})~\mbox{m}^{-3} \nonumber \\
& & \label{eqn:ionapprox}\\
\frac{n_{\rm i}^{2}}{n - n_{\rm i}} & = & A(T)
\end{eqnarray}
\citep{lang80} where we have tidied the RHS into the function $A(T)$.
Hence, we derive the quadratic
\begin{equation}
n_{\rm i}^{2} + A(T)\ n_{\rm i} - A(T)\ n = 0 
\end{equation}
and by restricting ourselves to the positive root arrive at
\begin{equation}
\iota = \frac{A}{2 n}\left(\sqrt{1+ \frac{4n}{A}} -1\right). 
\label{eqn:ionatphot}
\end{equation}

We see then that the ionization fraction $\iota$ and parameter $\beta_{\rm c}$
 are interdependant and should be solved for iteratively to ensure
consistency. Thus we can determine the lightcurve 
behaviour by initially calculating a value for $\beta_{\rm c}$ from 
(\ref{eqn:beta0def}) with a judiciously
chosen value for $\kappa_{1}$. Thereafter we use the number density from 
(\ref{eqn:rhoph}) to calculate the ionization fraction from 
(\ref{eqn:ionatphot}). This then allows us to find $\kappa_{1}$ from
 (\ref{eqn:kap1mixcomp}) and hence $\beta_{\rm c}$ again from 
(\ref{eqn:beta0def}) 
to better accuracy. We can repeat this iteratively to achieve the desired 
accuracy. This value can then be used to calculate the lightcurve 
behaviour at a given wavelelength.

\subsubsection{Mixed composition}

For the more realistic case of mixed composition, we cannot 
calculate the ionization state at a given position analytically. Instead we 
must iteratively solve the network of Saha equations under the prevailing
physical conditions to enable us to determine $\kappa_{1}$. This solution can 
then iterated with $\beta_{\rm c}$ around the equations (\ref{eqn:beta0def}), 
(\ref{eqn:rhoph}) and $\kappa_{1}$ loop in a similar way to the above. 

\subsubsection{Three Zone Model}

The above method works well in situations where either free-free or bound-free
opacity is dominating. Unfortunately, as we noted at the end of section
\ref{sec:derive}, in the situation of mixed opacity sources we can introduce
significant errors if we use a value for $\kappa_{1}$ assuming the incorrect 
source. We can rescue much of the above formalism, however, if we split the
spatial profile into 3 zones. We are almost bound to improve  our 
calculation regardless of where we place the boundaries of the zones but
clearly it makes sense  to try to ensure that we have free-free
dominated (outer regions), bound-free dominated (inner regions) and mixed 
(intermediate regions) zones. Assuming that hydrogen species provide the 
dominant  opacity source, we select the boundary between these regions by the 
hydrogen ionization fractions  $\iota=0.1$ and $\iota=0.9$ which occur at 
$\eta_{0.1}$ and $\eta_{0.9}$.

Specifically then, we need to replace the integral in 
equation~(\ref{eqn:spatint}) with
one over the three types of zone and rework equation~(\ref{eqn:beta0def}) to 
redefine $\beta_{\rm c}$. Calculating $x_{0.1}$ and $x_{0.9}$ using $y=0$ or 
$y_{\rm m}$ as desired, the integral in (\ref{eqn:spatint}) thus becomes
\begin{eqnarray}
& & \int_{-\infty}^{\infty} \kappa_{1}\ e^{-2\left(\frac{x}{a}\right)^{2}} 
\ {\rm d}\left(\frac{x}{a}\right)  \nonumber \\
& & =  \frac{2}{a}\int_{0}^{x_{0.1}} 
\kappa_{1,{\rm  bf}}\ e^{-2\left(\frac{x}{a}\right)^{2}}\ {\rm d}x  
 +   \frac{2}{a} \int_{x_{0.1}}^{x_{0.9}} \kappa_{1,{\rm m}} 
\ e^{-2\left(\frac{x}{a}\right)^{2}}\ {\rm d}x  
\nonumber\\ 
& & + \frac{2}{a}\int_{x_{0.9}}^{\infty} \kappa_{1,{\rm ff}} 
\ e^{-2\left(\frac{x}{a}\right)^{2}}\ {\rm d}x
\label{eqn:3zoneint} \\
& & = 2\kappa_{1,{\rm ff}}\ \mbox{erfc}
\left(\sqrt{2}\frac{x_{0.9}}{a}\right) \nonumber\\ 
& & + 2\kappa_{1,{\rm m}} 
\ \left[\mbox{erfc}\left(\sqrt{2}\frac{x_{0.1}}{a}\right) -
       \mbox{erfc}\left(\sqrt{2}\frac{x_{0.9}}{a}\right)\right] 
\nonumber\\
& & +  2\kappa_{1,{\rm bf}} 
\ \left[1 - \mbox{erfc}\left(\sqrt{2}\frac{x_{0.1}}{a}\right)\right]
\end{eqnarray}
which we use to replace a factor $\kappa_{1} \sqrt{\pi/2}$ in 
equation~(\ref{eqn:beta0def}).

The question remains of how to rapidly find values for $\eta_{0.1}$ and
$\eta_{0.9}$. Again, in the spirit of finding an easily calculable 
approximation, and noting that even inaccurately determining these boundaries 
will still improve our integral calculation, let us consider a situation where 
all species other than
hydrogen remain fully ionized. Using standard methods, we derive
\begin{equation}
\mu_{\rm e}=\frac{2}{1+X\ (2\iota-1)}.
\end{equation}
Incorporating this into equation (\ref{eqn:ionapprox}) along with our density 
profile, we can show
\begin{equation}
\rho_{0}\ e^{-\eta^{2}} = \frac{2\ m_{\rm H}}{\left[1+(2\ \iota-1)\ X\right]}
\frac{1-\iota}{\iota} A(T)
\end{equation}
which we can rearrange and solve directly for $\eta_{\iota}$. We note the 
particularly simple form this reduces to for $\iota=0.5$.

\begin{figure}
\includegraphics[angle=-90,scale=0.3]{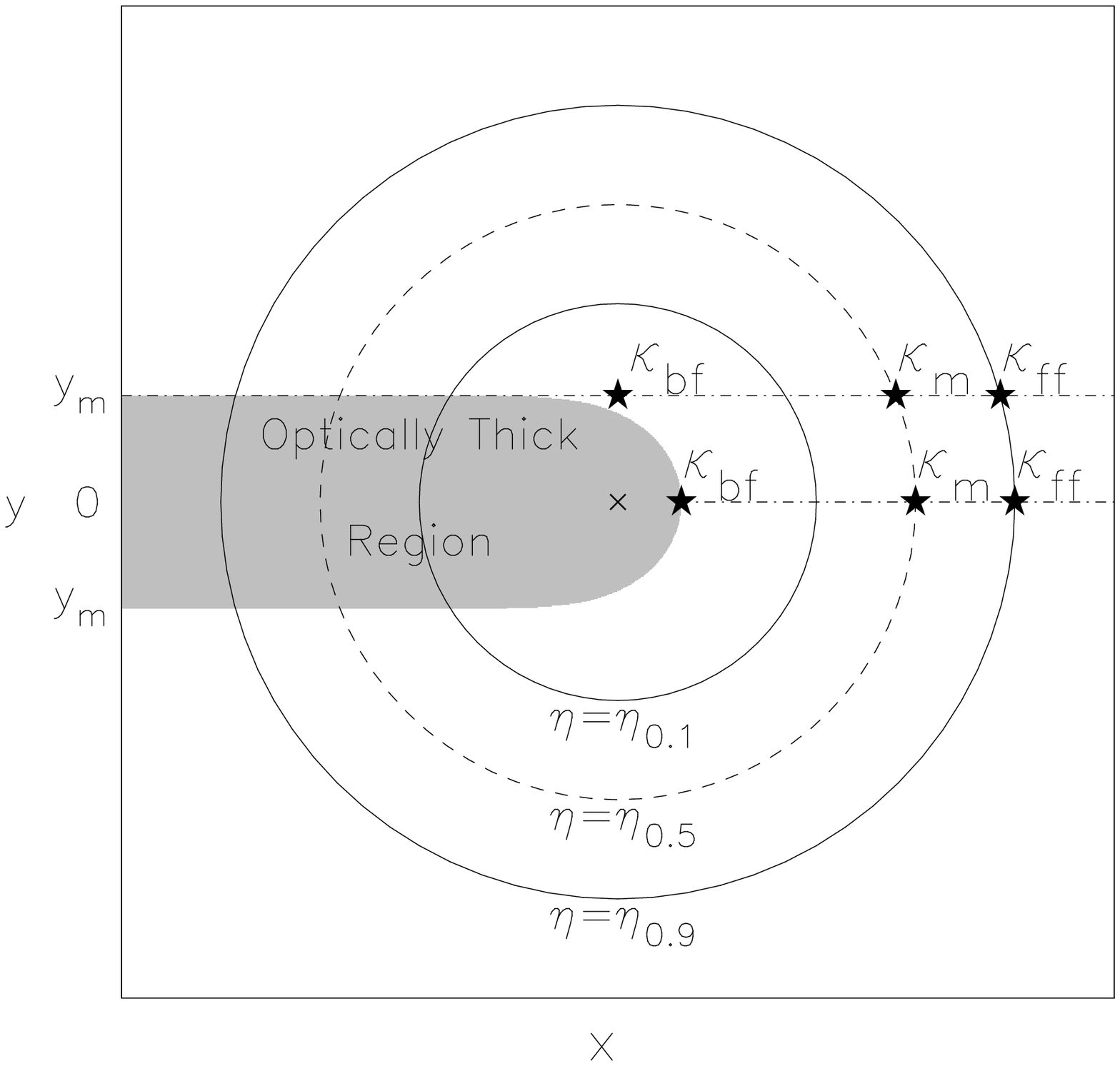}
\caption{Schematic representation of the alternative integration 
schemes. The dot-dashed lines mark the two possible lines of sight considered
for an observer (at $x=\infty$). Along the $y=0$ line of sight the one-zone
model would calculate $\beta_{\rm c}$ using $\kappa(=\kappa_{\rm bf}$) 
evaluated at the
photosphere. For a three zone scheme $\beta_{\rm c}$ would be calculated
using a combination of the relevant opacities evaluated at the indicated
positions. The boundaries for each zone occur at the intersection of the
dot-dashed line with the solid lines. Similarly the position where the
opacities would be calculated for an optically thick/thin transition line
of sight are indicated. While the circles of $\eta_{\iota}$ have been placed
purely for illustrative convenience, the position of the photosphere has
been calculated assuming $\tau_{0}=10$. 
Other possible combination exist depending on
the relative positions of the photosphere, peak density and ionization
boundaries.}
\label{fig:schemes}
\end{figure}

For similar reasons to the single zone model, we evaluate $\kappa_{\rm bf}$
and $\kappa_{\rm ff}$ at the point of highest density in their region 
(eg. either at $\eta_{0.9}$ or $\eta_{\rm max}$ for free-free). Given the 
rapid density variation in the mixed zone, we evaluate $\kappa_{\rm m}$ at 
$\eta_{0.5}$ or $\eta_{\rm max}$ as appropriate. The possible integration 
schemes are illustrated schematically in Figure~\ref{fig:schemes}.

\subsection{Frequency Dependence $\beta_{\rm c}=\beta_{\rm c}(\eta,\nu)$}

The parameter $\beta_{\rm c}$ has a direct frequency dependence from the 
$\nu^{-3}$ 
in opacity and also dependence through both $\kappa_{1}$ and $\epsilon$.
For regions where the Gaunt factors are slowly changing functions of frequency
eg. along the Paschen continuum in the $4000$--$8000$~\AA  ~range or when
free-free opacity is dominating, we ignore the 
small error introduced by neglecting the contribution of $g_{\rm ff}$ and 
$g_{\rm bf}$  to $\kappa_{1}$. Instead we need only
correct for the direct and $\epsilon$ frequency dependence of 
$\beta_{\rm c}$. Thus,
\begin{equation}
\beta_{\rm c}(\nu)\approx\beta_{\rm c}(\nu_{0}) 
\left(\frac{\nu_{0}}{\nu}\right)^{\frac{6}{10-\Gamma}} 
\left(\frac{\epsilon}{\epsilon_{0}}\right)^{\frac{2}{10-\Gamma}}. 
\label{eqn:nucorr}
\end{equation}
 With this correction we have only to calculate $\beta_{\rm c}$ accurately
at a single wavelength with the iterative method outlined above. The lightcurve
behaviour as a function of both wavelength and time then follows immediately 
from the combination of (\ref{eqn:nucorr}) and (\ref{eqn:fluxtot}).

\subsection{Time Dependence $\beta_{\rm c}=\beta_{\rm c}(\eta,\nu,\beta)$}

The dependence of $\beta_{\rm c}$ on time again comes through both 
$\kappa_{1}$ 
and $\epsilon$ (if $\Gamma\neq0$). In principle, we can use a similar
expression to equation~(\ref{eqn:nucorr}) also to correct for the $\epsilon$
time-dependence. However, the exponential nature of the expression
renders the form of the lightcurves more complex 
and less instructive than that derived above. As a result, 
this correction is probably best included only in numerical solutions. 
More seriously, we cannot, in general, predict the future 
ionization structure for  a mixed composition gas from its current state. Thus,
the time-dependence of $\kappa_{1}$ can, in general, only be included with 
numerical solution at a series of different times $\beta$. The exception to 
this rule is when we can be sure that the dominant opacity is and will remain
free-free throughout the time considered. 
In this case $\kappa_{1}$ is a constant in time and we can use the same rapid 
approximation as the previous section.

\subsection{Comparison of Integral Methods}
\label{sec:comparison}

We can lift the restriction to the purely free-free opacity
case by iterative numerical solution of the Saha equations for a gas of mixed 
composition at each time in the lightcurve. From the ionization profile
of each species we can in principle determine the opacity at any point.  

\begin{figure}
\includegraphics[angle=-90,scale=0.3]{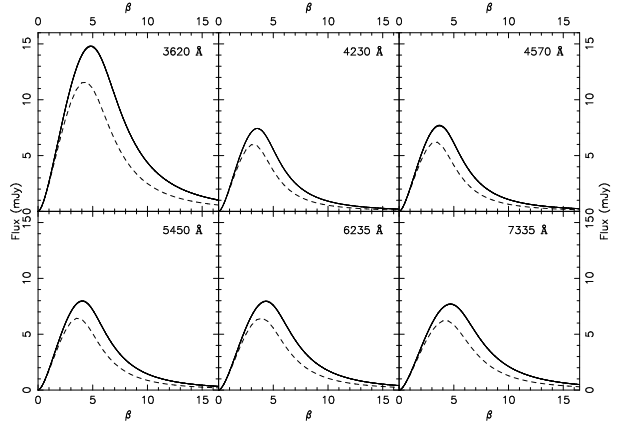}
\caption{Comparison of predicted lightcurves using the 4 combinations of
1- or 3-zone and $y=0$ or $y_{\rm m}$ line of sight integration methods.
The 4 methods produce  virtually indistinguishable results.
The parameters of the model are $M=3.0\times10^{16}$~kg, 
$a_{0}=7.5\times10^{6}$~m, $T_{0}=16~000$~K and $\Gamma=0$. A lightcurve 
calculated from a full numerical integration of the opacity through the 
fireball is plotted as a dashed line. A distance of 166~pc has been assumed.}
\label{fig:comp16k}
\end{figure}

\begin{figure}
\includegraphics[angle=-90,scale=0.3]{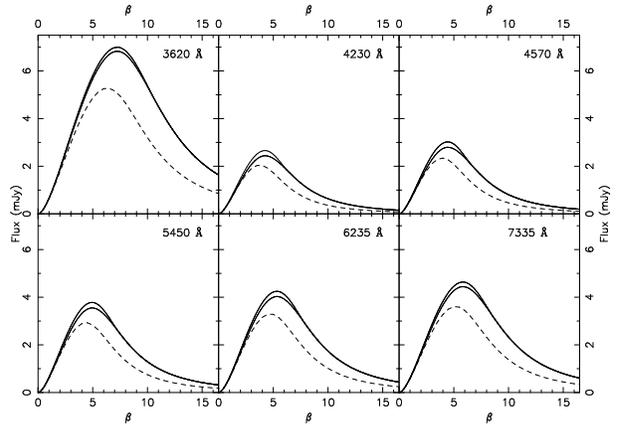}
\caption{The same comparison as Figure~\protect\ref{fig:comp16k} but with
$T_{0}=10~000$~K.The 1- and  3- zone models are indistinguishable in either 
case. The $y=0$ models produce slightly lower peak fluxes.}
\label{fig:comp10k}
\end{figure}

We compared the results achieved by the two integration lines of sight 
$y=y_{\rm m}$ or $0$ and using 1- or 3-zone integration schemes. 
The total continuum opacity was calculated numerically using 
using the methods of \citet{gray76} with the exception
of H$^{-}$ bound-free \citep{geltman62} and free-free \citep{stilley70}, 
He$^{-}$ \citep{mcdowell66} and HeI bound-free \citep{huang48}. Calculating
$\beta_{\rm c}$ from this opacity we can iterate to a consistent solution at
each time. Altering
the number of zones used yielded virtually no discernable effect on the 
predicted lightcurves. For temperatures around 16~000~K, the two lines
of sight considered produced results differing by at most around 0.5\%. 
However, at lower temperatures the predicted fluxes could differ more, 
reaching around 5\% at 10~000~K (see Figures~\ref{fig:comp16k} and  
\ref{fig:comp10k}).

\section{Comparison to Observation - Deducing Fireball Parameters}
\label{sec:fitting}

 The model 
was fitted to lightcurves for the flickering in the dwarf nova SS Cyg and 
the `bump phase' of SN 1987A. We employed a $\chi^{2}$ minimization
amoeba code \citep{numrec}
to derive the best-fit values for $M$, $a_{0}$, $T_{0}$, $t_{0}$, $H$, and 
$\Gamma$. We use the approximation that the ionization fractions
for all the species are well represented by the conditions at 
$x=0, y=y_{\rm m}$ and use the single zone integration approach
to arrive at a self-consistent solution for $\beta_{\rm c}$ and $\kappa_{1}$ 
there at each time.


\subsection{SS Cygni}

Rapid spectroscopy of SS~Cyg was carried out using the Low Resolution Imaging 
Spectrograph (LRIS: Oke et al. 1995) \nocite{oke95} on the 10-m Keck II 
telescope on Mauna Kea, Hawaii, between UT~09:38 and 09:56 on 6 July 1998, 
covering orbital phases 0.8276 to 0.8715. The instrumental set and data 
reduction were the same as
described by \citet{steeghs01}, \citet{obrien01} and \citet{skidmore03}. 
14,309 spectra were obtained
using the rapid data acquisition system. The spectra covered 3259-7985~\AA
~with 2.4~\AA~pixel$^{-1}$ dispersion and a mean integration time of
72.075~ms and no dead-time between individual spectra. Further details of 
these observations are given in \citet{skidmore04}.
A particular flickering event was isolated between UT~9:39:10 and UT~9:43:44 
and flux from other sources removed by fitting, for each wavelength, a low 
order polynomial to the fluxes both before and after the event.
Lightcurves were formed from the mean flux in the regions 3590-3650~\AA, 
4165-4270~\AA, 4520-4620~\AA, 5100-5800~\AA, 5970-6500~\AA ~and 7120-7550~\AA 
~at 2~s resolution. 

\begin{figure}
\includegraphics[angle=-90,scale=0.3]{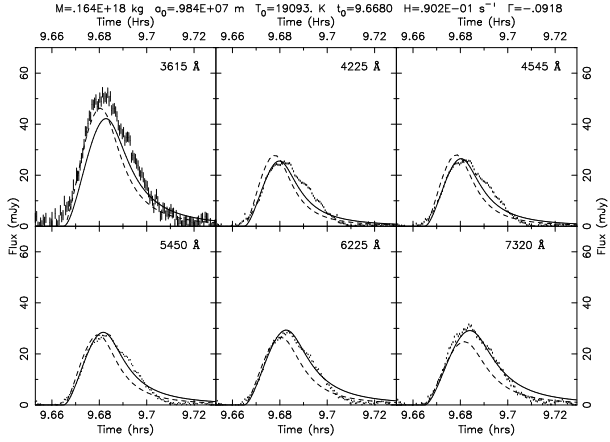}
\caption{Analytic fits (solid) to the SS Cyg data points and lightcurves
generated from numerically calculated spectra using the 
best-fit parameters (dashed). $\Gamma$ was allowed to be a fit parameter and 
the points were weighted according to the observational errors.}
\label{fig:SSlight1}
\end{figure}

\begin{figure}
\includegraphics[angle=-90,scale=0.3]{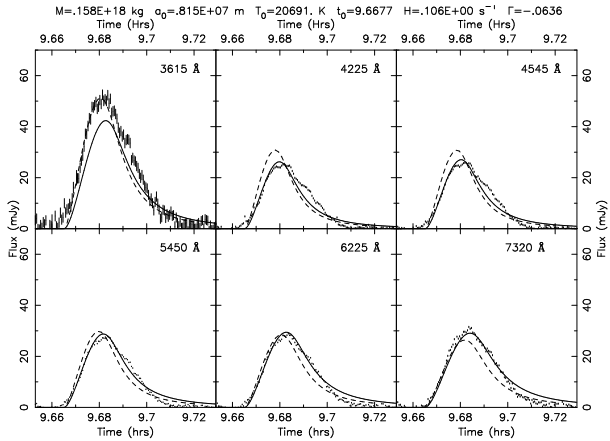}
\caption{Same comparison as Figure~\protect\ref{fig:SSlight1}, $\Gamma$ was 
allowed to be 
a fit parameter and the points were weighted equally.}
\label{fig:SSlight2}
\end{figure}

\begin{figure}
\includegraphics[angle=-90,scale=0.3]{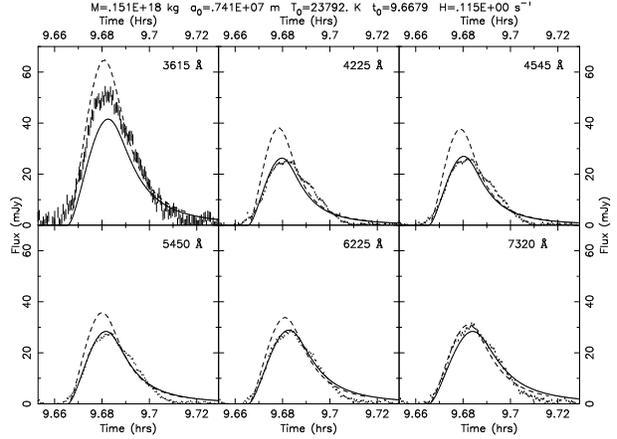}
\caption{Same comparison as Figure~\protect\ref{fig:SSlight1}, $\Gamma=0$ 
(isothermal) 
was fixed and the points were weighted according to the observational errors.}
\label{fig:SSlight3}
\end{figure}

\begin{figure}
\includegraphics[angle=-90,scale=0.3]{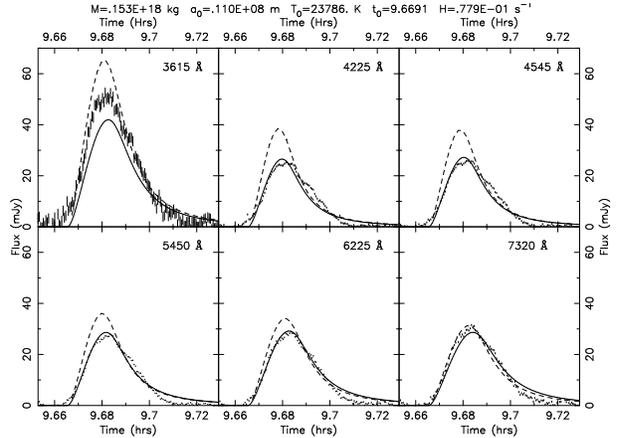}
\caption{Same comparison as Figure~\protect\ref{fig:SSlight1}, $\Gamma=0$ 
(isothermal) 
was fixed and the points were weighted equally.}
\label{fig:SSlight4}
\end{figure}

Model lightcurves were computed using a distance of  $166$~pc 
\citep{harrison99} and assuming solar composition. 
It appears that the disagreement is 
dominated by the systematic error of the oversimplicity of our model rather 
than observational errors. Accordingly, we 
carried out fits both weighting the data points according to their formal 
observational errors and with equal weighting.
We plot the derived analytic lightcurves alongside the
observations in Figures~\ref{fig:SSlight1} and \ref{fig:SSlight2}. 
From each set of derived parameters, we carried out a detailed numerical
integration of the opacity calculated with an LTE ionization varying across 
the density profile as described in \citet{pearson03}. These models 
allowed us to generate a timeseries of continuum spectra and ``numerical'' 
lightcurves 
for comparison to the data in each case. These numerical lightcurves are also
plotted in the Figures. The fitted value of
$\Gamma$ is close to zero in both cases and so we refitted the data fixing
$\Gamma=0$ exactly. These results are shown in Figures~\ref{fig:SSlight3} and
\ref{fig:SSlight4}. The best-fit parameters for each case are summarised in 
Table~\ref{tab:SSparams}.

Although we would expect the numerical lightcurves to more accurately represent
reality, the analytic lightcurves generally fit the data better. This is
unsuprising since the parameters used in both cases have been optimized for
the analytic forms. The numerical lightcurves do seem closer to the data in
the 3615~\AA window in Figures~\ref{fig:SSlight1} and \ref{fig:SSlight2}. However,
Figures~\ref{fig:SSlight3} and \ref{fig:SSlight4} and our other unpublished
lightcurves suggest that this is serendipitous. The difficulty in reproducing
the Balmer jump may well reflect that we have 5 fitting points on the Paschen
Continuum (wavelengths longer than 3646~\AA) and only one above the Balmer jump.
Since the continuum level will have a close to $\nu^{-3}$ relationship between
discontinuities, fixing the level at several points may well bias the parameters
towards an accurate fit here rather than the more complex interplay of 
parameters required to accurately reproduce the Balmer jump. In ideal 
circumstances we would like several more points at shorter wavelengths to 
address this issue.


\begin{table*}
\small
\begin{center}
\begin{tabular}{rcccccccc} \hline
Model & $M$  & $a_{0}$ & $T_{0}$ & $t_{0}$ & $H$ & $\Gamma$ & 
$v_{0}$ & $E_{\rm kin}$\\ 
 SS Cyg & ($10^{17}$ kg) &  ($10^{6}$ m) &  ($10^{4}$ K) & (UT) 
& (${\rm s}^{-1}$) & & (km ${\rm s}^{-1})$ & ($10^{28}$ J)\\
\hline
6 param. weight.   & 1.6 & 9.8 & 1.91 & 9.6680 & 0.090 & -0.092 & 890 
& 9.7 \\
6 param. unweight. & 1.6 & 8.1 & 2.07 & 9.6677 & 0.106 & -0.064 & 860
& 8.8 \\
Isothermal weight.    & 1.5 & 7.4 & 2.38 & 9.6679 & 0.115 & --- & 860 
& 8.3 \\
Isothermal unweight.  & 1.5 & 11.0 & 2.38 & 9.6691 & 0.078 & --- & 860
& 8.4 \\\hline 
SN 1987A & ($10^{31}$ kg) &  ($10^{13}$ m) &  ($10^{3}$ K) & (MJD) & 
($10^{-8}~{\rm s}^{-1}$) & &
(km ${\rm s}^{-1}) $ & ($10^{43}$ J)\\ \hline
5-band UBVRI & 2.6 & 2.1 & 4.72 & 47018.7 & $7.2$ & 0.066 & 1500 & 4.5 \\
4-band BVRI  & 2.5 & 2.2 & 4.69 & 47025.3 & $7.0$ & 0.071 & 1500 & 4.3 \\
\hline
\end{tabular}
\end{center}
\protect\label{tab:SSparams}
\caption{Derived and auxiliary parameters from the analytic fits to flickering
of SS~Cyg and the lightcurve of SN~1987A.}
\end{table*}

The derived masses, $M\sim1.6\times10^{17}~{\rm kg}$ are equivalent to 
$\sim400$~s of the estimated mean mass transfer
rate from the secondary of 
$\dot{M}\approx4\times10^{14}~{\rm kg}~{\rm s}^{-1}$ 
and $\sim2\times10^{-4}$ of the total disk mass 
$M_{\rm disk}\approx7\times10^{20}~{\rm kg}$ \citep{schreiber02}. Similarly,
the lengthscale of the expanding region, $a\sim9.1\times10^{6}~{\rm m}$ is 
$\sim$1.5\% of the estimated disk
radius $R_{\rm disk}\approx6\times10^{8}~{\rm m}$ \citep{schreiber02}. 
The kinetic energy is much greater than the thermal energy 
$E_{\rm th}\approx4\times10^{25} \left(\frac{M}{10^{17}~{\rm kg}}\right)
\left(\frac{T}{20~000~{\rm K}}\right)~{\rm J}$ but comparable to that of a 
putative, moderately strong, magnetic field in a sphere of radius $a_{0}$, 
$E_{\rm mag}=6\times10^{28} \left(\frac{a_{0}}{10^{7}~{\rm m}}\right)^{3} 
\left(\frac{B}{6~{\rm T}}\right)^{2}~{\rm J}$.

\begin{figure*}
\begin{minipage}{17cm}
\includegraphics[angle=-90,scale=0.65]{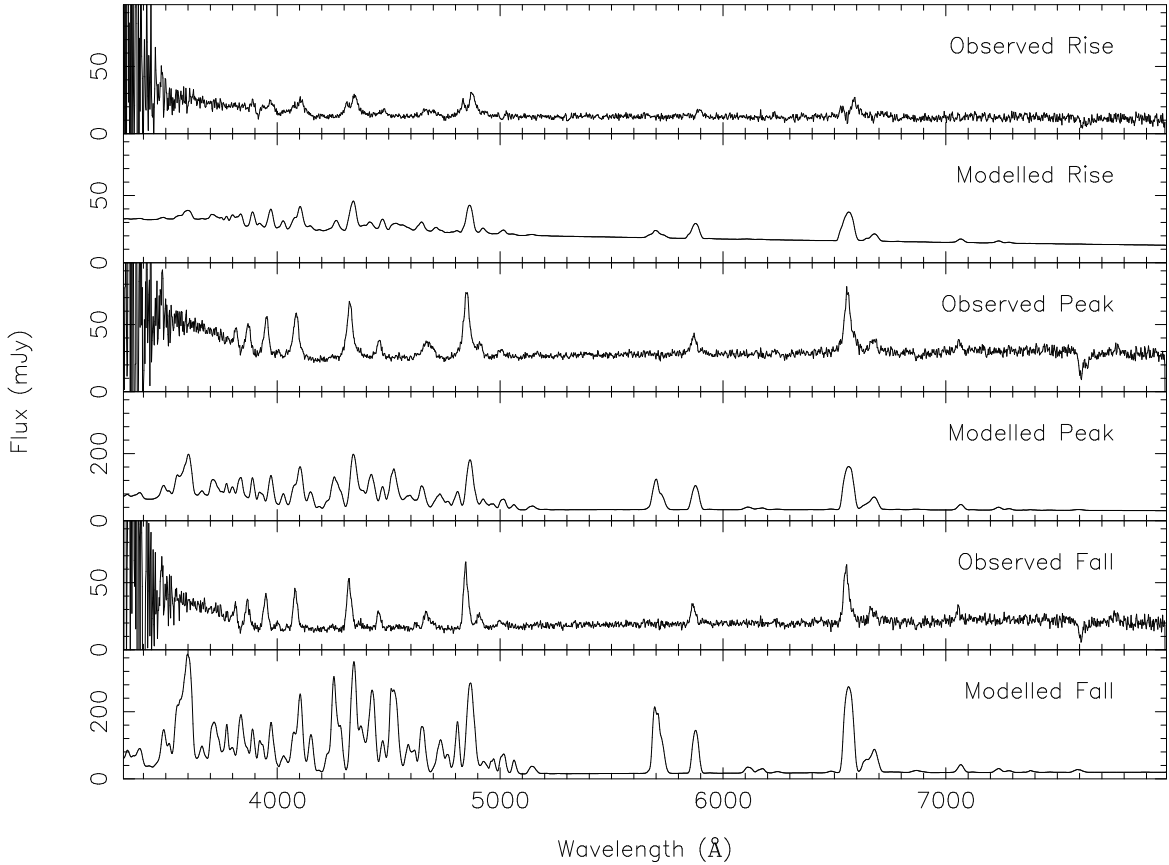}
\caption{Comparison of 3 spectra at UT~9.6722 (rise), 9.6830 (peak) and 
9.6916 (fall) calculated 
from numerical integration across our fireballs with the observed spectra of 
SS~Cyg. The `6 parameter weighted' values were used to generate the
numerical spectra.}
\label{fig:speccomp}
\end{minipage}
\end{figure*}

 We generated full optical spectra from the 
`6-parameter weighted' set of values and compare them to the observed spectra 
at three different times in Figure~\ref{fig:speccomp}. The model data have 
been convolved with the instrumental blurring of 9.8~\AA~FWHM. The results
show how parameters derived from the analytic forms for the continuum 
lightcurves can subsequently be used to reproduce spectra. The agreement
with observation at early times is remarkable. At the peak flux and later,
however, the models predict more metal and stronger lines than are observed, 
particularly in the $3000$--$5000$~\AA~ range. Many of the relative line 
strengths, however, are still reproduced.


\subsection{SN 1987A}

Given the similarities in the mechanisms underlying the flickering lightcurves
in our models and those of supernovae, we attempted to fit the 
UBVRI photometric 
data for SN~1987A obtained from \citet{catchpole87}. The data were converted 
to fluxes using the standard values and effective 
wavelengths for the bands in \citet{enca&a}.
The fits assumed an interstellar reddening E(B-V)=0.15 
\citep{hamuy88,fitzpatrick88} and assumed the
same distance to the LMC of $50.1$~kpc as \citet{catchpole87}. 
Given the results of sophisticated NLTE spectral and lightcurve 
modelling (eg. Mitchell et al. 2002) \nocite{mitchell02} we should not expect
an accurate match between our simple models and the SN~1987A observations. 
However, it will allow us to highlight the underlying similarities of these 
problems and to gauge the robustness of our derived parameters. 
The fits used only the data after JD~2446875.0 to eliminate the flash
phase and the data were all assigned equal weights. For the sake of continuity 
with the SS~Cyg fit,
we continued to allow 6 fit parameters rather than fixing
the combination of $t_{0}$ and $H$ to give a launch time of JD~2446849.82 
when the associated neutrino event was detected by Kamiokande-II 
\citep{koshiba87}. Since we have not attempted to remove a ``background'' level
the fit may be contaminated by the early stages of the radioactive tail. 
However, the flux in this region is dominated by the expanding/collapsing 
photosphere effect and any heating should show up in the 
fitted value of $\Gamma$ differing from zero.


\begin{figure}
\includegraphics[angle=-90,scale=0.3]{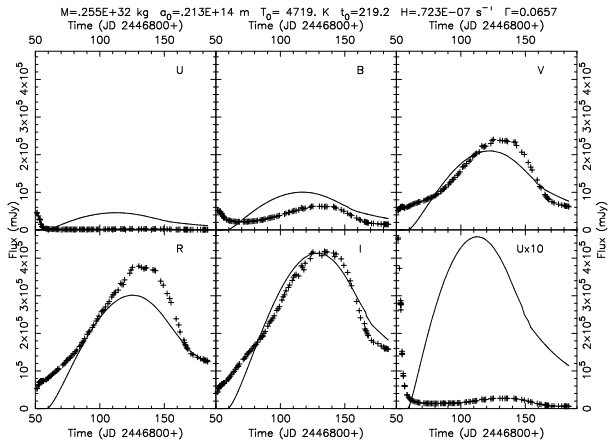}
\caption{Analytic fits (solid) to the UBVRI SN~1987A data of 
\protect\citet{catchpole87}. $\Gamma$ was allowed to be a fit parameter and
all the datapoints with $t>{\rm JD}~2446875.0$ were given equal weight.}
\label{fig:SNlight}
\end{figure}
\begin{figure}
\includegraphics[angle=-90,scale=0.3]{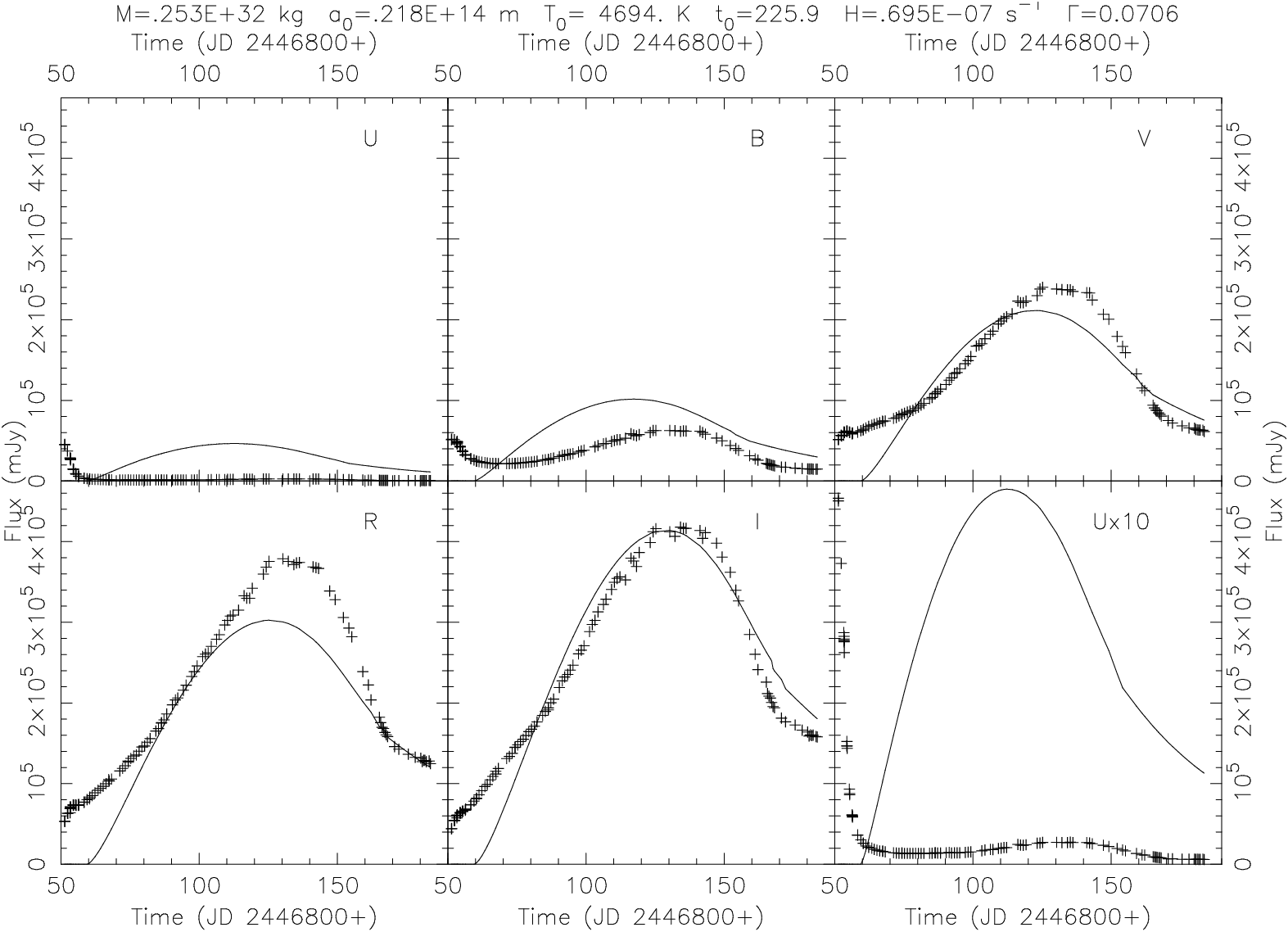}
\caption{Same comparison as Figure~\protect\ref{fig:SNlight} fitting only
BVRI band data.}
\label{fig:SNlight2}
\end{figure}


We see in Figure~\ref{fig:SNlight} that we can achieve a qualitative match
between the analytic expression and the data. However, the  U~band is 
particularly poorly reproduced, representing the effect of blanketing
noted by \citet{menzies87}. We refitted to the data using just the BVRI~bands
and show the results in Figure~\ref{fig:SNlight2}. The parameters for
the fits are included in Table~\ref{tab:SSparams}. We see that eliminating
the U~band data has relatively little effect and derive a value for
the mass of the material in the expansion of 
$M=2.6\times10^{31}~{\rm kg}~(\approx13M_{\odot})$. This compares to
an estimate of $7M_{\odot}<M_{\rm env}<11M_{\odot}$ by \citet{saio88}. The
typical expansion velocity of $1500~{\rm km}~{\rm s}^{-1}$ compares to a value
derived from measurements of the blueshifts in P~Cygni profiles of
$\sim2500~{\rm km}~{\rm s}^{-1}$ at maximum light \citep{mccray93}. Typical 
estimates of
$E_{\rm kin}\sim2\times10^{44}~{\rm J}$ \citep{bethe90,woosley88} are somewhat 
larger than our $E_{\rm kin}\approx4\times10^{43}~{\rm J}$.

\begin{figure}
\includegraphics[angle=-90,scale=0.3]{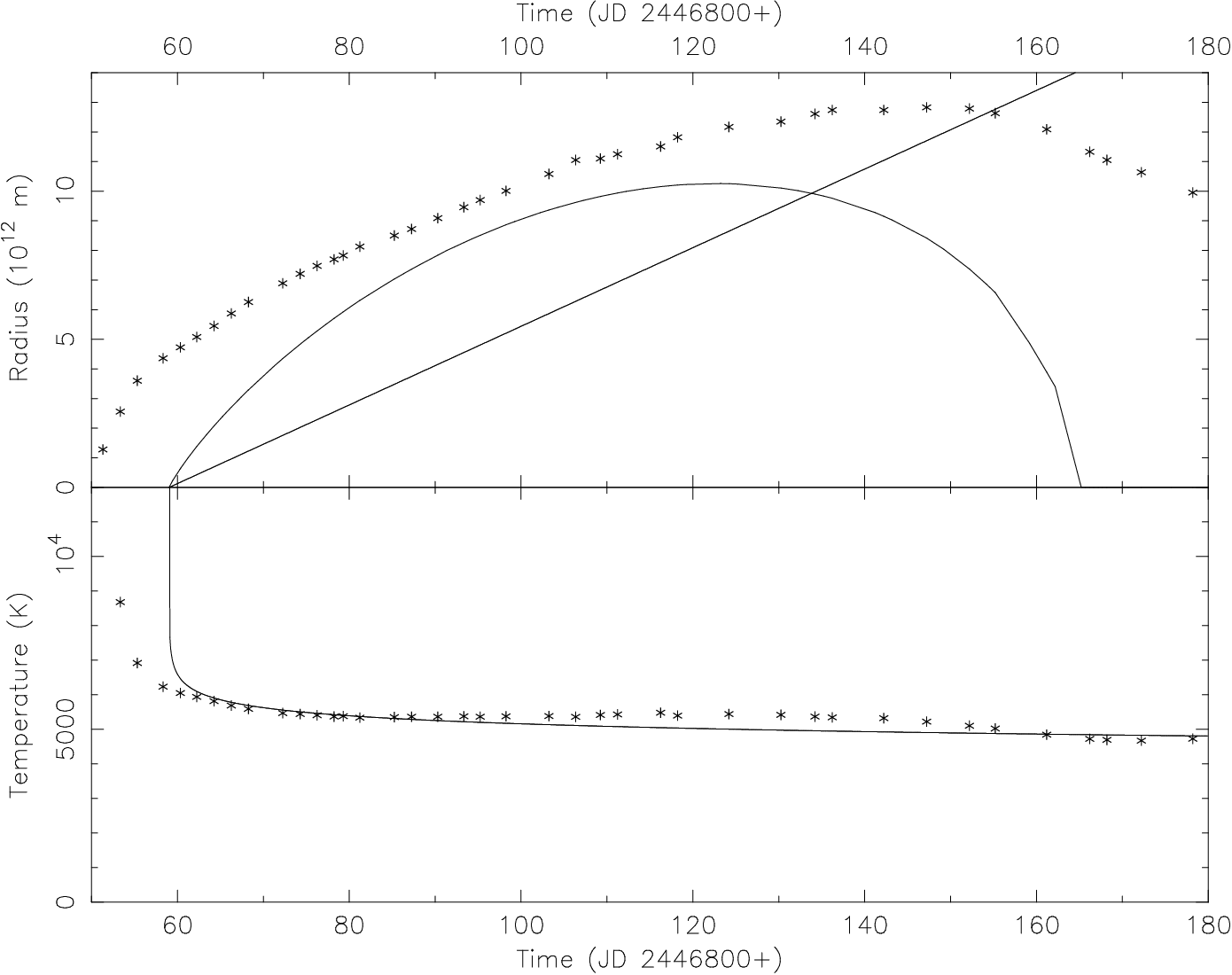}
\caption{The temperature and radius derived by \protect\citet{catchpole87}
(asterisks) and temperature, optically thick radius (in the V band) and 
current lengthscale from our UBVRI fit parameters.}
\label{fig:SNcomp}
\end{figure}

It is unsuprising that 
the lightcurves do not match in detail, since the Gaussian density profile
we have assumed does not accurately represent supernova ejecta. In particular,
 the reversal of the fast rise and slow decline
derived analytically can be understood from the way in which a photosphere
marching inwards through a shell (in comoving coordinates) evolves. In the 
shell case, the material will first become optically thin at low impact 
parameters once the photosphere reaches the  inner surface of 
the shell and later at higher impact parameters with their
greater path-lengths through the shell material.  While these can not 
be considered in any sense a good fits, we do retrieve values for the mass, 
size and temperature (see Figure~\ref{fig:SNcomp}) for this phase
comparable with those given by 
\citet{catchpole87}. We also see the same isothermal behaviour from the 
lightcurve analysis as those authors. The size of our photosphere is 
less than the radius
they derive which is consistent with our including both optically thick and 
thin emission regions whereas they treat all the emission as optically thick.  
Our lightcurves also sometimes show an intriguing feature 
(unfortunately most apparent in the poorly fitted U band) of a 
point of inflexion at around JD~2446960. This is
reminiscent, in timing and in character, of the break in the lightcurves 
attributed to heating
by radioactive decay. In our lightcurves, however, it appears to arise from 
the point at
which the material becomes completely optically thin with no contribution
from an optically thick core region. 
All these factors, showing reasonable results for SN~1987A despite the 
inherent simplicity of our model, reinforce our confidence in the robustness 
of the parameters derived for SS Cyg which had much better agreement between 
data and theory.


\section{Summary}

We have derived a formalism and analytic expressions applicable to a variety 
of systems that reproduces the evolution of their optical lightcurves. The 
method has been tested for two widely different cases allows us
to derive reasonable values for the physical parameters involved in the
expansion. There is encouraging agreement with a method using a full 
integration of the opacity across the expanding region to generate both 
lightcurves and spectra. We have seen how the
flickering of close  binary systems can be modelled as smaller, hotter 
analogues of the well known rebrightening ``bump'' in supernovae lightcurves.
We hope to test this  approach further by comparison to data from LMXBs and 
other systems in the future.

\acknowledgments{We thank the referee for the thoughtful and helpful suggestions in 
response to an earlier version of this paper. KJP has been supported, in 
part, by the U.S. National Science Foundation through grant AST-9987344 and, 
in part, through LSU's Center for Applied Information Technology and Learning.
He also thanks Juhan Frank for stimulating discussions that improved
this paper.}

\end{document}